\documentclass[journal,draftcls,onecolumn,12pt,twoside]{IEEEtranTCOM}
\normalsize

\usepackage{cite}
\usepackage{amsmath,amssymb,amsfonts}
\usepackage{algorithm}
\usepackage{algorithmic}
\usepackage{graphicx}
\usepackage{textcomp}
\usepackage{xcolor}
\usepackage{multicol}
\usepackage{multirow}
\usepackage{subcaption}
\usepackage{mathrsfs}
\usepackage{mathrsfs}
\usepackage{soul}
\usepackage{setspace}

\newcommand{\NR}[1]{\textcolor{orange}{#1}}
\newcommand{\FR}[1]{{\color{blue} FR: #1}}
\newcommand{\KV}[1]{{\color{magenta} KV: #1}}

\usepackage{lastpage}
\usepackage{graphics}
\usepackage{wrapfig,lipsum,booktabs}
\usepackage{comment}
\usepackage{mathtools}

\begin{document}


\title{CSI-Based Data-driven Localization Frameworking using Small-scale Training Datasets in Single-site MIMO Systems}


\author{Katarina Vuckovic, Farzam Hejazi, and Nazanin Rahnavard \vspace{-1mm}}

\maketitle



\begin{abstract}

This work presents a date-driven user localization framework for single-site massive Multiple-Input-Multiple-Output (MIMO) systems. The framework is trained on a geo-tagged Channel State Information (CSI) dataset. Unlike the state-of-the-art Convolutional Neural Network (CNN) models, which require large training datasets to perform well, our method is specifically designed to operate with small-scale training datasets. This makes our approach more practical for real-world scenarios, where collecting a large amount of data can be challenging. 
Our proposed FC-AE-GPR framework combines two components: a Fully-Connected Auto-Encoder (FC-AE) and a Gaussian Process Regression (GPR) model. 
Our results show that the GPR model outperforms the CNN model when presented with small training datasets. However, the training complexity of GPR models can become an issue when the input sample size is large. To address this, we propose using the FC-AE to reduce the sample size by encoding the CSI before training the GPR model.
Although the FC-AE model may require a larger training dataset initially, we demonstrate that the FC-AE is scenario independent. This means that it can be utilized in new and unseen scenarios without prior retraining. Therefore, adapting the FC-AE-GPR model to a new scenario requires only retraining the GPR model with a small training dataset. 

\end{abstract}

\section{Introduction}
\label{Intro}
The abundance of smartphone devices and wireless networks has paved a new era in the Location-Based Services (LBS) applications.
Nowadays, LBSs are widespread across many different domains including navigation systems, emergency  services, travel planning, asset management, location-based recommendation, and geo-social networking \cite{10.1145/3292390.3292394}. Although Global Positioning System (GPS) currently represents a ubiquitous outdoor positioning system, its limited accuracy (to 5 m for civilian applications) and its degraded performance in rich-scattering environments (i.e. urban canyons and indoor environments) \cite{8226757}, make it a non-ideal solution {for} positioning applications that require higher accuracy. Examples of such applications include autonomous driving, indoor asset and people localization and augmented reality applications. To this end, researchers have been investigating leveraging the wireless communication systems for localization purposes. 

{Sub-6 GHz and millimeter-Wave (mmWave) frequency Multiple-Input Multiple-Output (MIMO) antenna configurations are at the core of the 5G wireless communication technology.}
In addition to high data rates and increased spectral efficiency, these technologies also provide high temporal resolution \cite{8755880}. Some studies have been exploring these properties by using Angle-of-Arrival (AoA) and Time-of-Arrival (ToA) to propose geometry-based solutions \cite{vuckovic2021mapcsi,9013365}. The advantage of geometry-based models is that they do not require a large dataset to train the model. However, their performance typically falls short when compared to deep learning neural network models. Thus far, the highest localization accuracy has been achieved by applying the fingerprint localization method to deep Convolutional Neural Networks (CNNs) \cite{sun2019fingerprint,hejazi2021dyloc}.
\textit{Fingerprinting}, in the context of localization, is a technique that maps every location in the environment to a unique wireless measurement during the offline training phase  and then uses this geo-tagged map to make location estimations based on measurements in the online phase. The key idea exploited in the fingerprinting localization is that wireless channel between the user and the Base Station (BS) is uniquely determined by the scattering environment surrounding the user's location \cite{8307353}. 
One commonly used parameter is the Receive Single Strength (RSS). Since the RSS provides a single measurement from the BS or Access Point (AP),  multiple APs are required to generate a unique fingerprint. Therefore, RSS fingerprinting is used in wireless system that have rich AP distributions such as Wireless Sensor Networks (WSNs) \cite{9440373,mehrareview}, WiFi networks \cite{KUMAR20161,9647082,zhang2019wireless,xie2020analyzing}, or Distributed Massive MIMO (DM-MIMO) systems \cite{7390953,9217356}. 

The second type of fingerprinting parameters takes advantage of the multipath characteristics such as Channel State Information (CSI). Unlike the RSS parameters that require multiple APs, CSI fingerprints can be used for single-site localization but require a MIMO antenna to generate a unique fingerprint. Instead of using CSI in its raw format, some studies convert the CSI matrix into an Angle Delay Profile (ADP) matrix by applying a linear transformation~\cite{hejazi2021dyloc}. The ADP combines the information of the Power Delay Profile (PDP) with the AoA information in MIMO systems and creates semantic visual interpretation of the channel multipaths. 
Several fingerprinting techniques based on CSI or ADP data have been proposed for massive MIMO systems \cite{vieira2017deep, sun2019fingerprint,9128640, ferrand2020dnn, de2020mamimo, hejazi2021dyloc}. The best performance is observed by training a CNN with ADP fingerprints \cite{hejazi2021dyloc}. However, the drawback of using a data-driven approach such as CNN is that it requires collecting measurements for the entire environment. Performing such an elaborate data collection task is not only time consuming but also impractical for real world applications. 
Additionally, any changes in the existing scenario call for continuous model re-training and updated datasets. Hence, the inability to generalize on unseen environments creates a major hurdle for practical implementation of the model. These challenges make it difficult for models to be employed in real world scenarios.  


{To alleviate the data collection burden, researchers are looking towards solutions that can generalize well on smaller datasets. Some studies have been exploring transfer learning \cite{nguyen2021transfer} techniques to adapt their models to new environments  \cite{9129126,9860906, guo2022transfer} . 
Due to the small number of hyper-parameters, Gaussian Process Regression (GPR) models have shown the ability to train models with small scale datasets. The GPR model has already been used for RSS-based fingerprinting for Indoor Positioning Systems (IPS) with multiple APs \cite{KUMAR20161,6061737,7275083,liu2016gaussian,zhao2018gaussian, zhang2019wireless,xie2020analyzing}. Some of these works have investigated the impact of reducing the size of the training dataset but none of them compare the performance to CNN. DeepMap \cite{9097850} proposes using Deep Gaussian Processes Regression (DGPR) \cite{damianou2013deep} to construct an indoor radio map and perform-RSS based fingerprinting. The authors design a two-layer DGPR and compare it to a single layer GPR. The simulation results show that DeepMap significantly improves the performance of GPR when the dataset is small. DGPR is again used in \cite{teng2018localization} to improve the indoor localization reliability using RSS measurements. However, the only work that tackles single-site outdoor localization using GP models is in \cite{9414388} where the authors propose a Deep Convolutional Gaussian Process (DCGP) regression model \cite{blomqvist2018deep} for mmWave Massive MIMO outdoor localization. The simulation results show that the DCGP regression outperforms CNN regression fingerprinting using ADPs. Even though, DCGP demonstrates improved accuracy performance over CNN, the dataset {size} required 
to train the CNN and DCGP is the same, authors' do not discuss training with datasets of different sizes. The "deep convolution" part of the DCGP still requires a larger training dataset and the true benefit of GPR models (ability to train on small-scale datasets) is not exploited.  

The most prominent weakness of the standard GPR training is that it suffers from high computational and memory complexity. The complexity of the model often poses a limitation in many applications. Specifically the GPR has training computational complexity of $O(n^3)$ and memory complexity of $O(n^2)$, where $n$ is the number of training points in the dataset  \cite{GPmeetsBigData}. 
{Additionally, the computational complexity is also impacted by 
the dimensionality ($d$) of the input data sample vector}.  Although, some kernels can be more computationally efficient than others, the computational complexity can quickly increase as the size of the input data sample vector grows.
Therefore, the computational time can be reduced by either using a small training dataset or by reducing the dimension of the input data sample vector. 
When designing for massive MIMO systems, the size of the ADP is directly proportional to the number of array antennas and subcarriers. Therefore, this paper proposes an Auto-Encoder (AE) to reduce the size of the ADP matrix prior to training a GPR model. The AE is designed to be independent of the scenario and requires only one time training after which it can be tested (deployed) on scenarios with similar ADP fingerprints. 

The contributions of this paper can be summarized as follows:
\begin{itemize}
   \item We argue that the current state-of-the-art deep learning models pose a major challenge in real world implementation due to burdensome task of extensive data collection and the limited abilities to generalize beyond the training dataset. Therefore, we propose a novel architecture based on GPR models that requires a much smaller dataset to train. We compare the performance of the our architecture to the CNN and demonstrate that our model outperforms the CNN for small-scale training datasets. 
   \item Next we discuss that the memory requirement and computational complexity of the GPR model pose a challenge as the size of the input sample increases (ADP becomes larger). Therefore, we propose a Fully-Connected Auto-Encoder (FC-AE) to reduce the dimension of the input prior to training the GPR model. Even though the FC-AE may require a larger dataset to train, we argue that the FC-AE is independent of the scenario. Therefore, the training is performed only one time and then the FC-AE can be applied to new, unseen scenario with similar ADP signatures. 
    \item{We have developed a new model, FC-AE-GPR, which combines the FC-AE and GPR models. The trained FC-AE model can be transferred to new scenarios without any prior retraining, while the GPR model has to be retrained for new scenarios but requires only a small training dataset. This allows our model to implicitly perform transfer learning to tackle new scenarios.
    }
    \item Finally, we present complex GPR optimization model to find the optimal kernel type and its associated hyper-parameters.
    Specifically, the optimization model considers five different kernel types and performs hyper-parameter optimization for each of them. Then, the best kernel type and hyper-parameter combination is selected. 
\end{itemize}

 The rest of the paper is organized as follows.
 The channel model and the ADP matrix are defined in Section \ref{Model}. Following a comprehensive GPR model discussion in Section \ref{GPR}, the localization framework is introduced in Section \ref{Framework}. The simulation dataset and training model results are explained Section \ref{Simulations}. The localization performance is examined in Section \ref{Results}.  In Section \ref{Conclusion}, the paper is concluded by summarizing the work and highlighting the main points.

\section{Channel Model}
\label{Model}
\subsection{Channel Model}
\label{Chan Model}
Consider a  typical MIMO Orthogonal Frequency-Division Multiplexing (OFDM) wireless network with a single BS. Similar to \cite{ali2017millimeter}, assume the BS is equipped with a Uniform Linear Array (ULA) antenna  with $N_t$ antenna elements and it uses OFDM signaling with $N_c$ subcarriers. Furthermore, the users' equipment has a single omni-directional antenna.
The channel between the BS and the user is modeled using COST 2100 \cite{liu2012cost} with $C$ distinguishable clusters. Moreover, each cluster constitutes $R_C$ distinguishable paths. Each path can be characterized by a delay $\tau_{m}^{(k)}, k \in \{ 1, \dots,C\}, m \in \{ 1, \dots,R_C\}$, an angle of arrival to the BS's antenna $\theta_{m}^{(k)}$ and a complex gain $\alpha_{m}^{(k)}$ \cite{ali2017millimeter}.
Given a wide-band OFDM system, $\tau_{m}^{(k)} = n_{m}^{(k)} T_s$, where $T_s$ and $n_{m}^{(k)}$ denote the sampling duration and the sampled delay belonging to the path $m$ of the cluster $k$, respectively \cite{sun2019fingerprint}.  Then the bandwidth of each subcarrier is $f = {1}/{N_cT_s}$ and $f_l = lf$ is the $l^{th}$ subcarrrier. 
Channel Frequency Response (CFR) for each subcarrier $l$ can be written as \cite{alkhateeb2016frequency}
\vspace{-3mm}
\begin{equation}
\small{
    \boldsymbol{h}[l] = \sum_{k=1}^{C} \sum_{m=1}^{R_C} \alpha_{m}^{(k)} \boldsymbol{e}(\theta_{m}^{(k)}) e^{-j 2\pi \frac{l \: n_{m}^{(k)}}{N_c} }
    \label{eq_CSIdef}
    \vspace{-3mm}
    }
\end{equation}
\normalsize
where $\boldsymbol{e}(\theta) = [1,e^{-j2 \pi \frac{d cos(\theta)}{\lambda}},\dots,e^{-j2 \pi \frac{(N_t - 1)d cos(\theta)}{\lambda}}]^T\,$ denotes the array response vector of the ULA and $d$ is the gap between two adjacent antennas 
The overall CFR matrix of the channel between the BS and the user can be expressed as $ \boldsymbol{H} = [\boldsymbol{h}[1],\dots,\boldsymbol{h}[N_c]]$. This matrix is commonly known as CSI.

\subsection{Angle-Delay Profile}
\label{ADP}

The ADP matrix is a linear transformation of the CSI matrix computed by multiplying the CSI matrix with two Discrete Fourier Transform (DFT) matrices \cite{vuckovic2021mapcsi}. The transformation maps the space frequency domain CSI to the angle and delay domain \cite{sun2019fingerprint}. Referring to \cite{hejazi2021dyloc}, the DFT matrix $\boldsymbol{V} \in \mathbb{C}^{N_t \times N_t}$ is defined as 
\begin{equation}
   [\boldsymbol{V}]_{\: z,q} \overset{\Delta}{=} \frac{1}{\sqrt{N_t}} e^{-j 2 \pi \frac{\left(z(q-\frac{N_t}{2})\right)}{N_t} },
   \label{eq_V}
\end{equation}
and $\boldsymbol{F} \in \mathbb{C}^{N_c \times N_c}$ as
\begin{equation}
   [\boldsymbol{F}]_{\: z,q} \overset{\Delta}{=} \frac{1}{\sqrt{N_c}} e^{-j 2 \pi \frac{zq}{N_c} }.
   \label{eq_F}
\end{equation}
Then, the ADP matrix $\boldsymbol{A}\in \mathbb{R}^{N_t \times N_c}$ is defined as 
\begin{equation}
   \boldsymbol{A} =| \boldsymbol{V}^H \boldsymbol{H} \boldsymbol{F}|, 
   \label{eq_ADP}
   \vspace{-5mm}
\end{equation}
where $|\cdot|$ is the absolute value.

The ADP has several unique characteristics that make it the preferred input over the raw CSI information. Unlike the CSI that has no semantic interpretation, the ADP is a visual representation of all distinguishable paths between the user and the BS \cite{hejazi2021dyloc}. 
The ADP matrix example in Fig. \ref{fig_ADP} has two strong multipath components at $(q,z) = (5,1)$ and $(q,z)= (5,15)$ which correspond to angle and delay values of $(\theta, \tau)=(80 ^{\circ}, 0.1$ ns)  and $(\theta, \tau)=(80 ^{\circ},1.5$ ns), respectively. The range of the top x-axis is from 0 to $N_t - 1$ and the range of the right y-axis is from 0 to $N_c-1$, as shown on the top x-axis and right y-axis, respectively.
Referring to \eqref{eq_V}, the matrix $\boldsymbol{V}$ is a phase shifted DFT that maps the antenna domain CSI to angle domain such that each element in the matrix indicates the channel gain corresponding to the multipath angles \cite{8307353}. Furthermore, the matrix $\boldsymbol{F}$ in \eqref{eq_F} maps the matrix to delay domain. Thus, $|A|_{q,z}$ is interpreted as the power of the path associated with the angle  $\theta_q = \arccos{(\frac{2q-N_t}{N_t})}$ and delay $\tau_z = zT_s$, where $q=0,\dots,N_t-1$ and $\tau_z = zT_s, z=0,\dots,N_C-1$. The angle and delay corresponding to $N_t$ and $N_c$ are represented on the bottom x-axis and the left y-axis, respectively. 
 \begin{figure}[h]
 \vspace{-5mm} 
    \centering
    \includegraphics[width=3.4 in,height=2.4 in]{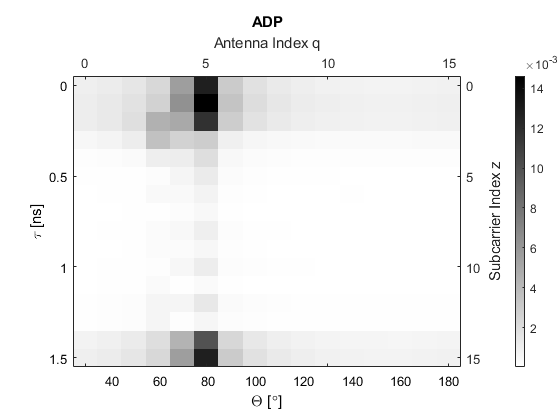}
    \caption {\small { Example of $N_t \times N_c = 16 \times 16$ ADP matrix. The top x-axis and right y-axis show the antenna index $q$ and subcarrier index $z$, respectively. The bottom x-axis and left y-axis show the angle $\theta$ and delay $\tau$, respectively.} }
   \vspace{-10mm} 
    \label{fig_ADP}
\end{figure}

Furthermore, the transformation in \eqref{eq_ADP} converts the dense CSI matrix into a sparse ADP representation. It is easier to compress a sparse matrix using encoding methods and smaller inputs reduce the complexity of the GPR optimization problem. Finally, referring to \cite{vieira2017deep}, ADP is highly correlated in location domain and similarities between the ADPs decrease smoothly with respect to their physical distances. This last property is important for the GPR model as it is used for the prior assumption in the Bayesian model. 
\section{Gaussian Process Regression} 
\label{GPR}

\subsection{Motivation for using Gaussian Processes}
The GPR is defined as a non-parametric supervised machine learning model based on Bayesian statistics \cite{wang2021intuitive}. 
Non-parametric algorithms do not make any assumptions regarding the form of the function that is to be estimated. This makes it a versatile model capable of estimating a wide variety of functions. The CNN models have also shown remarkable abilities to model complex functions. However, CNN models typically require large datasets to train the model effectively while the GPR model have shown the better generalization abilities on smaller training datasets \cite{9097850}. 
Furthermore, from a Bayesian perspective, a prior over the parameters is defined. The prior is an assumption that is encoded about the parameters before introducing the observation\cite{GPforML}. By selecting the kernels of the GPR model, the assumption about the spatial relationship between the input points is embedded into the model. 

\subsection{Gaussian Processes}
\label{GPRModel}
A \textit{Gaussian Process} (GP) is a collection of random variable functions indexed by time or space. The key property of a GP is that any finite subset of the random variables is jointly Gaussian distributed. For any finite set of vector elements $\mathbf{x}_1,...,\mathbf{x}_n \in\mathcal{X}$, the associated set of random variables $f(\mathbf{x}_1),...,f(\mathbf{x}_n)$ follows a joint Gaussian distribution.
This joint distribution is fully determined by the mean vector $\boldsymbol{\mu}$ and covariance matrix $\mathbf{K}$, where $\boldsymbol{\mu}=(m(\mathbf{x}_1),...,m(\mathbf{x}_n))^T$ and $[\mathbf{K}]{ij}=k(\mathbf{x}_i,\mathbf{x}_j)$, such that

\begin{small}
\begin{equation}
    \begin{bsmallmatrix}
 f(\mathbf{x}_1)\\
 \vdots \\
 f(\mathbf{x}_n)
\end{bsmallmatrix} 
\sim \mathcal{N} \left( \begin{bsmallmatrix}
 m(\mathbf{x}_1)\\
 \vdots \\
 m(\mathbf{x}_n)
\end{bsmallmatrix},
\begin{bsmallmatrix}
k(\mathbf{x}_1,\mathbf{x}_1) & \dots & k(\mathbf{x}_1,\mathbf{x}_n) \\
\vdots & \ddots & \vdots \\
k(\mathbf{x}_n,\mathbf{x}_1) & \dots & k(\mathbf{x}_n,\mathbf{x}_n) 
\end{bsmallmatrix}
\right).
\label{eq_joint}
\end{equation}
\end{small}
The following notation is commonly used in literature to represent the GP
 \begin{equation}
    f(\mathbf{x})\sim \mathcal{GP}(m(\mathbf{x}),k(\mathbf{x},\mathbf{x'})),
\label{eq_GP}
\end{equation}
where the mean and covariance functions are defined as
 \begin{equation}
    m(\mathbf{x}) = \mathbb{E}[ f(\mathbf{x})],
\label{eq_mean}
\vspace{-5mm}
\end{equation}
 \begin{equation}
   k(\mathbf{x},\mathbf{x'}) = \mathbb{E}[ (f(\mathbf{x})-m(\mathbf{x}))(f(\mathbf{x'})-m(\mathbf{x'})]
\label{eq_cov}
\end{equation}
for any $\mathbf{x},\mathbf{x'} \in \mathcal{X}$ \cite{7275083}.  
Without any prior knowledge of the mean, the default assumed value is zero \cite{chen2021recent}. After estimation, if it is determined that the predicted values are offset by some mean value, the mean can be added back after the prediction step. This is called \textit{centering} the function. 
On the other hand, the covariance matrix not only describes the shape of our distribution, but ultimately determines the characteristics of the function that we want to predict \cite{gortler2019visual}. Therefore, the focus of the GPR model design is on the covariance function and its associated hyper-parameters.

\subsection{Covariance Functions}
In statistics, the covariance measures the correlation between two random variables $\mathbf{x}_i$ and $\mathbf{x}_j$ \cite{rice2006mathematical}, while the covariance function (or \textit{kernel}) describes the spatial or temporal correlation of a random process. Therefore, given a set of input points $\{\mathbf{x}_i | i = 1,...,n\}$, we can construct the covariance matrix $\mathbf{K}$
 whose entries are the kernel functions {$k(\mathbf{x}_i, \mathbf{x}_j)$} \cite{GPforML}. The covariance matrix is also known as the Gram matrix. 

The kernel defines the relationship between the points, thus encoding the assumptions about the function that we want to learn \cite{GPforML}. The basic assumption is that near by points are highly correlated. Therefore, training points that are near the testing point, should be informative about the prediction at that test point. In the case of localization, the ADP fingerprints represent the input points for which we wish to define the spatial correlation using the kernel function. A kernel should be selected such that the correlations between the ADP inputs is high for two close by locations and monotonically decreases as the distance between two locations increases. 
There are multiple kernels that satisfy this criteria. A list of common monotonically decreasing kernels is shown in Table \ref{tb_covfun},  in which $\sigma$ is the standard deviation of the signal, $l$ is the characteristic kernel scale,  $\alpha$ is the  positive-valued scale-mixture parameter, and
\begin{equation}
\vspace{-4mm}
\mathbf{r} = \sqrt{(\mathbf{x}_i -\mathbf{x}_j)^T(\mathbf{x}_i -\mathbf{x}_j)}.
\label{eq_cov_r}
\end{equation}

\begin{table}[h]
\vspace{-5mm}
 \begin{center}
\caption{\small{Several Covariance Functions with Monotonically Decreasing Kernels.}
}
\scalebox{0.9}{%
\begin{tabular}{ |c|c|c|} 

 \hline
\textbf{Name} & \textbf{ Kernel $k(\mathbf{x}_i,\mathbf{x}_j$)} &\textbf{Hyper-parameters $\Theta$} \\
 \hline
    Squared Exponential & $\sigma^2 exp(-\frac{\mathbf{r}^{2}}{2l^2})$ & ($\sigma,l$) \\
     \hline
 Exponential & $\sigma^2 exp(-\frac{\mathbf{r}}{2l^2})$ &  ($\sigma,l$) \\
  \hline
 Rational Quadratic & $\sigma^2 (1+\frac{\mathbf{r}}{2\alpha l^2}){-\alpha}$  & ($\sigma, \alpha, l$) \\
 \hline
 Matern32 & $\sigma^2 (1+\frac{\sqrt{3}\mathbf{r}}{ l})exp(-\frac{\sqrt{3}\mathbf{r}}{l})$  & ($\sigma, l$) \\
 \hline
 Matern52 & $\sigma^2 (1+\frac{\sqrt{5}\mathbf{r}}{ l} + \frac{5\mathbf{r}^{2}}{ 3l^2})exp(-\frac{\sqrt{5}\mathbf{r}}{ l})$  & ($\sigma, l$) \\
\hline
\end{tabular}}
 \label{tb_covfun}
 \vspace{-15mm}
\end{center}  
\end{table}

\subsection{Gaussian Process Regression Model}
{In the GPR model, the GP is used to model a continuous function by assuming that any point in the function's domain is drawn from a multivariate normal distribution \cite{GPforML}.}  
Let $\mathcal{D}_{train} \overset{\Delta}{=}(\mathbf{X,y})\overset{\Delta}{=}\{\mathbf{x}_i, y_i\}_{i=1}^n, \mathbf{x}_i \in \mathbb{R}^d, y_i \in \mathbb{R}$ be an input-output pair training dataset. Furthermore, assume that a latent function $f(\cdot)$ is responsible for generating the observed output $y_i$ given the input vector $\mathbf{x}_i$. Then, GPR model can be defined as 
\begin{equation}
    y_i = f(\mathbf{x}_i) + \epsilon_i,
\label{eq_gpr}
\end{equation}
where $f(\mathbf{x})\sim \mathcal{GP}(m(\mathbf{x}),k(\mathbf{x},\mathbf{x'}))$ 
, $\epsilon \sim \mathcal{N}(0, \sigma^2 \mathbf{I})$ is the noise of the system that has an independent, identically distributed (i.i.d.) Gaussian distribution with zero mean and variance $\sigma^2$, and $i$ refers to the $i^{th}$ observation. 

Given the training dataset $\mathcal{D}_{train}$ and a testing dataset $\mathcal{D}_{test} \overset{\Delta}{=}(\mathbf{X_*},\mathbf{y_*})\overset{\Delta}{=}\{\mathbf{x_*}_i,{y_*}_i \}_{i=1}^n, \mathbf{x_*}_i \in \mathbb{R}^d, {y_*}_i \in \mathbb{R}$, the posterior predictive distribution is obtained from the conditioning properties of jointly Gaussian distribution in \eqref{eq_joint}. The joint distribution of the observed values $\mathbf{y}$ and the predicted function $\mathbf{y_*}$ at the new testing sample $\mathbf{X_*}$ is defined as

\begin{equation}
    \begin{bmatrix}
 \mathbf{y}\\
 \mathbf{y_*}
\end{bmatrix} 
\sim \mathcal{N} \left(0,\begin{bmatrix}
\mathbf{K}+\sigma_n^2I & \mathbf{K_*}\\
\mathbf{K_*}^T & \mathbf{K_{**}}
\end{bmatrix}
\right),
\label{eq_post}
\end{equation}
where  $\mathbf{K} = K(\mathbf{X,X})$, $\mathbf{K_*}=K(\mathbf{X,X_*})$, and $\mathbf{K_{**}}=K(\mathbf{X_*,X_*})$ \cite{wang2021intuitive}.  
Then, using the rules of conditional distribution, it follows that 
\begin{equation}
    \mathbf{y_*}|(\mathbf{y,X,X_*})\sim\mathcal{N}(\mu_*,\Sigma_*)
\label{eq_gp_1}
\end{equation}
where the mean $\overline{\mathbf{y}}_*$ and variance $\mathbb{V} [ \overline{\mathbf{y}}_*]$ of the unknown function $\mathbf{y}_*$ are computed as \cite{Zhang2010,wang2021intuitive}
\begin{equation}
    \mu_*=\overline{\mathbf{y}}_{*} = \mathbf{K_*^{T}}(K+\sigma_n^2I)^{-1}\mathbf{y},
\label{eq_mu}
\end{equation}
\begin{equation}
     \Sigma_* = \mathbb{V} [ \overline{\mathbf{y}}_{*}] = \mathbf{K_{**} }- \mathbf{K_*^{T}}(K+\sigma_n^2I)^{-1}\mathbf{K_*}.
\label{eq_sigma}
\end{equation}

\subsection{Hyper-parameter Optimization}

The hyper-parameters describe properties of the kernel and the noise in the GP. Bayesian optimization is used to find the optimal hyper-parameters given by the vector $\mathbf{\Theta}$ as defined in Table \ref{tb_covfun}. The hyper-parameters can be estimated by minimizing the negative log marginal likelihood (NLML) function \cite{GPforML}
\begin{small}
\begin{align}
\label{eq_nlml}
   \log ({P}(\mathbf{y|X,\Theta}))= 
     -\underbrace{\frac{\mathbf{y}^T(\mathbf{K}+\sigma_n^2I)^{-1}\mathbf{y}}{2}}_{\textrm{model-fit}} 
   -\underbrace{\frac{\log |\mathbf{K}+\sigma_n^2I|}{2}}_{\textrm{complexity penalty}}  
   -\underbrace{\frac{n  \log(2\pi)}{2}}_{\textrm{normalization constant}}.
\end{align}
\end{small}

The NLML in \eqref{eq_nlml} has three distinctive terms that can be readily interpreted. The first is the \emph{model-fit} term, which is the only term that includes the observed targets. The second is the \emph{complexity penalty} term, which only depends on the covariance function and the inputs. Finally, the last term is the \emph{normalization constant}. The hyper-parameters are optimized by computing the partial derivative of the NLML with respect to $\mathbf{\Theta}$ and then using a gradient descent method to update the hyper-parameters in the GPR model at every iteration. 

\section{Localization Framework}
\label{Framework}

The localization framework consists of a Fully-Connected Auto-Encoder (FC-AE) model that first encodes the ADP matrix reducing its size. This is followed by a GPR model that predicts the user's location using the encoded ADP. The localization framework that combines the FC-AE with the GPR model is referred to as the FC-AE-GPR model. 

The diagram in Fig. \ref{fig_framework} illustrates the three phases of the FC-AE-GPR model. There are two offline training phases and an online testing phase. The first offline phase is only performed one time, while the second offline phase is repeated for every new environment. Even though the second offline phase is scenario dependent, it only requires a small training dataset to adapt to the new environment. 
In the first offline phase, the FC-AE is trained with the dataset from Scenario I. The FC-AE requires a larger dataset to train the model but the model is only trained one time. Therefore, all the data from Scenario I is used to train the FC-AE. 
In the second offline phase, the GPR is trained on a small training dataset from Scenario II. The small training dataset is 10\% or less of the entire Scenario II dataset. Before training the GPR model, the input ADP sample is first encoded using the FC-AE trained in the previous phase. Then, the GPR is trained on the encoded ADPs. 
Finally, in the online phase, the FC-AE-GPR model is tested on the remaining 90\% of Scenario II data.

  \begin{figure}[h]
    \centering
    \includegraphics[width=5.2in,height=2.8 in]{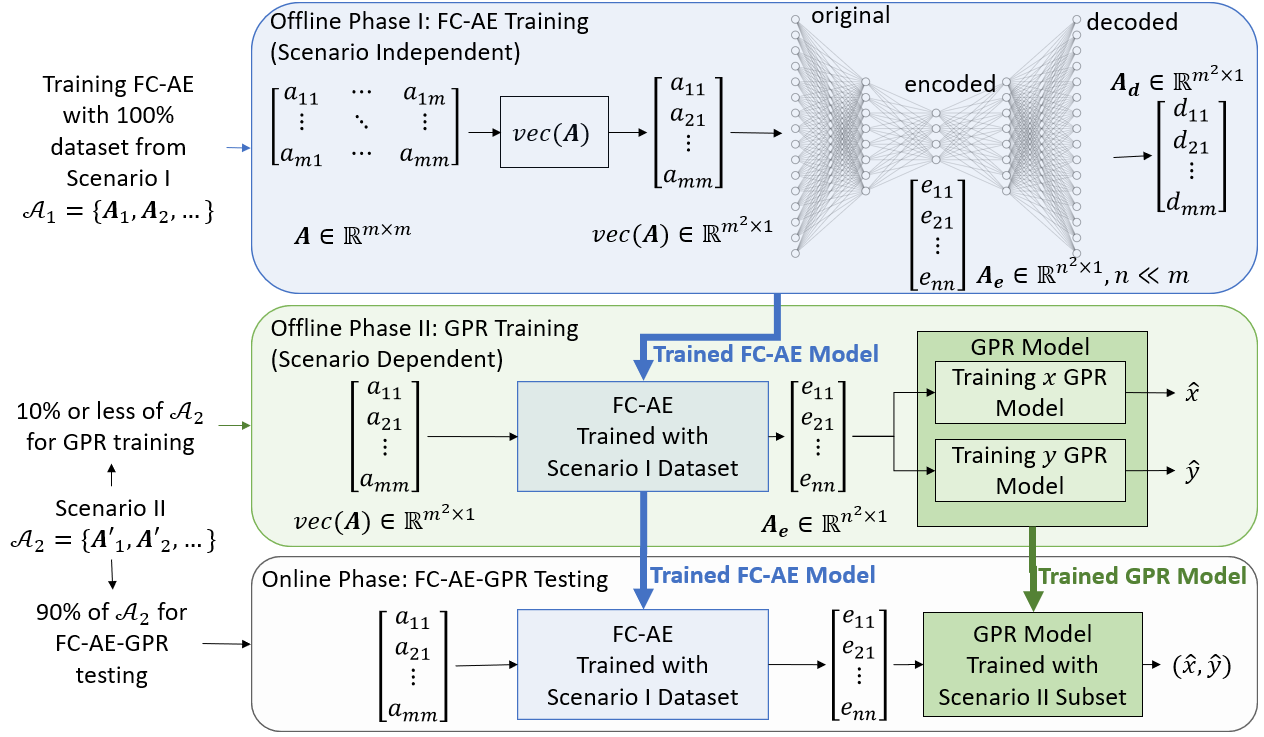}
    \caption {\small {Our introduced localization framework showing the two offline phases for training the FC-AE and GPR models and the online phase that tests the combined FC-AE-GPR model.}}  
    \vspace{-10mm}
    \label{fig_framework}
\end{figure}

\subsection{Fully-Connected Auto-Encoder (FC-AE)}

The FC-AE has an architecture similar to a Multi-Layer Perceptron (MLP) \cite{GA}.
An example of the FC-AE is illustrated in the first offline phase in Fig. \ref{fig_framework}. In this example, there is one hidden layer between the input layer and the encoded layer. The number of perceptrons in each layer decreases until it reaches a "bottle neck" (encoded layer). This part of the architecture is known as the encoder that compresses the input into latent representation.  The second part of the architecture is the decoder which is symmetrical to the encoder such that the number of perceptrons is increased at every layer. The output of the decoder has the same number of perceptrons as the input of the encoder. The objective of the decoder is to reconstruct the original input from the latent representation.

The principle behind training {an} FC-AE is that the AE is trying to reconstruct (decode) the input. Therefore, the model contains a reconstruction loss that penalizes the model when the reconstruction is different from the inputs. The similarity is often 
computed as the normalized correlation $\mathscr{S}$ between the input $\boldsymbol{A}$ and the decoded output $\boldsymbol{A_d}$ as
\begin{equation}
   \mathscr{S}(\boldsymbol{A},\boldsymbol{A_d}) = \frac{\mathrm{vec}(\boldsymbol{A}).\mathrm{vec}(\boldsymbol{A_d})}{||\boldsymbol{A}||_F||\boldsymbol{A_d}||_F},
   \label{eq_sim}
\end{equation}
where vec($.$) denotes an operator that concatenates columns of a matrix into a vector, operation $.$ denotes inner product and $||.||_F$ denotes the Frobenius norm. 

\subsection{Offline Phase I: FC-AE Training}

The size  of the ADP in \eqref{eq_ADP} depends on the number of antenna array elements ($N_t$) and the number of subcarriers ($N_c$) such that $\boldsymbol{A}\in\mathbb{R}^{N_t\times N_c}$. When $N_t$ and $N_c$ are small, the size of the ADP is computationally manageable for GPR model. However, as the number of antenna elements or the number of subcarriers increases, the ADP input becomes quite large and computationally exorbitant. The FC-AE is introduced to reduce the size of the ADP prior to training the GPR model. The objective of decreasing the size of the ADP input is to reduce the workload of the GPR model.

Parameters $N_t$ and $N_c$ are assumed to be equal $N_t=N_c=n$ to simplify the model. However, this is not required and can be extended to any  $N_t$ and $N_c$ size. As may be observed in the first phase in Fig. \ref{fig_framework}, the ADP $\boldsymbol{A}\in\mathbb{R}^{n\times n}$ is first converted to a vector $\mathrm{vec}(\boldsymbol{A})\in\mathbb{R}^{n^2\times 1}$. 
The objective of the FC-AE is to first compress the ADP such that the encoded ADP is $\boldsymbol{A_e}\in\mathbb{R}^{m^2\times 1}$, where $m<<n$, and then decode the ADP back to $\boldsymbol{A_d}\in\mathbb{R}^{n^2\times 1}$ such that there is minimal loss between the original $\mathrm{vec}(\boldsymbol{A})$ and the decoded $\boldsymbol{A_d}$. 

The FC-AE is a deep neural network that requires a large dataset to train the encoder and decoder. The FC-AE is trained with the dataset from Scenario I. For a given BS in Scenario I and an area covered by this BS, ADP samples at every location are collected to generate the dataset. In practice, the dataset for training the FC-AE does not have to be collected in only one scenario. 
Unlike the GPR that correlates the ADP to a user's location which is specific to the scenario, the FC-AE only aims to compress the ADP presentation into a lower dimensional space. Therefore, the FC-AE can be trained with data from any scenario or with a dataset comprised of samples from multiple scenarios. The dataset may even be a synthetically generated dataset. The more diverse the training dataset, the more scenarios can use the FC-AE model. 
This makes the FC-AE independent of the scenario, which enables the FC-AE-GPR model to be easily adapted to new, unseen scenarios. 


\subsection{Offline Phase II: GPR Training}
The second offline phase is 
designed to train the GPR model. The first step is to use the FC-AE trained in the first offline phase to encode the ADPs. Then, the encoded ADPs are used to train the GPR model. Referring to the GPR model defined in \eqref{eq_gpr}, the input to the GPR model is the encoded ADP and the output is the estimated user's location. Since the GPR model in \eqref{eq_gpr} is a single output model, two models are trained in parallel to estimate the $\hat{x}$ and $\hat{y}$ coordinates. 

To demonstrate the independence of the FC-AE with respect to the scenario, the dataset used to train the GPR models is collected from a new scenario which is labeled as Scenario II in Fig. \ref{fig_framework}.  
The two GPR models are trained using a small subset of the encoded ADP samples in Scenario II. The entire dataset in Scenario II contains an ADP measurement for every location in the environment. However, to train the GPR models, only 10\% (or less) of the samples in the dataset are used. The remaining 90\% of samples are used for testing. 
Unlike the FC-AE that is environment independent,  training the GPR model is required for every new scenario as it relates the ADP to the physical location of the user. Therefore, the ability to train the GPR model with a small training dataset is emphasized. 
\subsection{Online Phase: Testing Framework}
Finally, the online testing phase combines the trained FC-AE model with the trained GPR model. 
The online testing phase is performed on the remaining 90\% of the Scenario II dataset that was not used for the GPR training. In the first step the vectorized ADP is encoded using the FC-AE model. To validate that the encoded ADP can be trusted, the {normalized correlation} in \eqref{eq_sim} is used to compute the similarity between the original and decoded ADP. If the similarity is above a set threshold $thresh$ the FC-AE can be trusted, otherwise the FC-AE has not been trained with the appropriate ADP representative and cannot be used to estimate the user's location. If the ADP passes the threshold criteria, then encoded ADP is presented to the GPR models that estimate the user coordinates  $\hat{x}$ and $\hat{y}$ using \eqref{eq_mu}. This is summarized in Algorithm \ref{alg}.

\begin{algorithm}[h]
\algsetup{
linenosize=\scriptsize,
linenodelimiter=:
}
\caption{Online Phase: User Localization Testing}
\begin{algorithmic}[1]

\label{alg1}
\footnotesize
\REQUIRE{measured CSI at time $t$  ($\boldsymbol{H}$}); similarity threshold $thresh$  \\

\STATE{Convert $\boldsymbol{H}$ to ADP $\boldsymbol{A}$ using \eqref{eq_ADP}}  
\vspace{2pt}
\STATE{Reshape ADP into vector $\boldsymbol{A}_v\in\mathbb{R}^{n^2\times 1}$ $\leftarrow$ $\mathrm{vec}(\boldsymbol{A}\in\mathbb{R}^{n\times n})$ }  
\vspace{2pt}
\STATE{$\boldsymbol{A_e}\in\mathbb{R}^{m^2\times 1}$  $\leftarrow$  $\mathrm{encoder}(\boldsymbol{A}\in\mathbb{R}^{n\times n})$, $m<n$}
\vspace{2pt}
\STATE{$\boldsymbol{A_d}\in\mathbb{R}^{n^2\times 1}$  $\leftarrow$  $\mathrm{decoder}(\boldsymbol{A}\in\mathbb{R}^{m^2\times 1})$}
\vspace{2pt}
\STATE{$s$ $\leftarrow$ $\mathscr{S}(\boldsymbol{A},\boldsymbol{A_d})$}
\IF{$s > thresh$}
    \STATE{$\hat{x}$ $\leftarrow$ $\mathrm{GPR_x}(\boldsymbol{A_e})$}
    \STATE{$\hat{y}$ $\leftarrow$ $\mathrm{GPR_y}(\boldsymbol{A_e})$}
\ELSE
    \STATE{Encoded ADP ($\boldsymbol{A_e}$) is not valid}
\ENDIF
\vspace{2pt}
    \end{algorithmic}
    \label{alg}
\end{algorithm}
\vspace{-5mm}

\section{Simulations}
\label{Simulations}

\subsection{Simulation Dataset}
\label{Dataset}
The dataset of CSI and location pairs is generated using the DeepMIMO dataset framework \cite{Alkhateeb2019}.
DeepMIMO is a public dataset generation framework that allows the user to select from multiple environments and to define the parameters of the channel models and the BS antenna. DeepMIMO channels are
constructed based on accurate ray-tracing data. The parameters used to generate the dataset are summarized in Table \ref{tb_dataset}. 
The selected outdoor ($O1$) scenario is an urban environment with two streets and one intersection (Fig.~\ref{fig:scenario}). 
To be consistent with the single-site channel model, only one BS is active with a ULA antenna with $N_t$ antenna elements aligned in the y-axis and there is a single user equipped with an omni-directional antenna.
{Furthermore, the bandwidth is set to 100 MHz and OFDM signaling with $N_c$ number of subcarriers is employed.}
The number of paths is set to maximum which is 25.

\begin{table}[h]
\vspace{-5mm}
 \begin{center}
\caption{\small{Dataset Parameters}}

\begin{tabular}{ |c|c| } 
 \hline
Environment & Outdoor 1 ($O1$)\\ 
 \hline
 Frequency Band & 3.5 GHz\\ 
  \hline
  Bandwidth & 100 MHz\\ 
  \hline
  Base Station Antenna Orientation & ULA aligned in y-axis \\ 
 \hline
  Antenna Spacing & 1/2 wavelength \\ 
 \hline
  Antenna Elements ($N_t$) & 4,8,16  \\
  \hline
  Subcarrier Number ($N_c$) & 4,8,16 \\
  \hline
  Max Number of Paths & 25  \\
\hline
\end{tabular}
\vspace{-10mm}
 \label{tb_dataset}
\end{center}  
\end{table} 

\begin{figure}[h]
 \vspace{0 mm}
    \centering
    \includegraphics[width=3.3in,height=2.2 in]{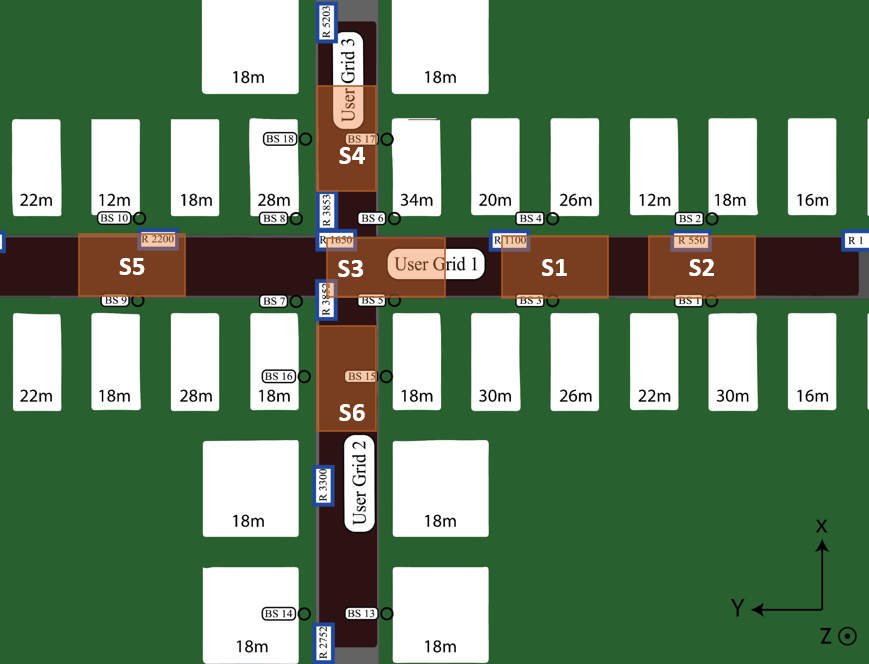}
    \caption {\small {DeepMIMO dataset top view of the outdoor environments $O1$ showing six scenarios S1 to S6. }}
    \vspace{-5mm}
    \label{fig:scenario}
\end{figure}

 The outdoor environment $O1$ is shown in Fig.  \ref{fig:scenario}. Six different scenarios are considered in this environment. The scenarios are labeled S1 to S6 and their properties are listed in Table \ref{tb:scenario}. One BS is active per scenario and a total of 72,400 samples are collected for each scenario. The rows are used to indicate the start and end position of the scenario. The separation between two adjacent locations is 20 cm for S1-S5 and 10 cm for S6. The area occupied by S1-S3 and  S5 is 36 m $\times$ 80 m, S4 is 135 m $\times$ 5.3 m, and S6 is 22 m $\times$ 13.2 m.
The different scenarios are used to demonstrate the independence of the FC-AE with respect to the scenario.

\begin{table}[h]
\vspace{-5 mm}
 \begin{center}
\caption{\small{Scenario Parameters}}

\begin{tabular}{ |c|c|c|c| } 
 \hline
 Scenario & Active BS & Rows & Area \\ 
  \hline
S1 & BS4 & R800-1200 & 36 m $\times$ 80 m\\
 \hline
S2 & BS1 & R300-700 & 36 m $\times$ 80 m\\
 \hline
S3 & BS5 & R1300-1700 & 36 m $\times$ 80 m\\
 \hline
S4 & BS17 & R4000-4400 & 135 m $\times$ 5.3 m\\
 \hline
S5 & BS10 & R2000-2400 & 36 m $\times$ 80 m\\
 \hline
S6 & BS15 & R3400-3800 & 22 m $\times$ 13.2 m\\
\hline
\end{tabular}
\vspace{-10mm}
 \label{tb:scenario}
\end{center}  
\end{table} 
Dataset S1 is used to train the FC-AE, then testing is performed on the remaining five scenarios S2-S6. In addition, to demonstrating that the FC-AE improves with large and more diverse  dataset, the FC-AE is retrained with a dataset consisting of both S1 and S4 data points, and tested with the remaining four scenario datasets. 
On the other hand, the GPR and CNN localization models are trained for each scenario separately. However, for each scenario only a small subset of the samples is used for training while the remaining samples are used for testing. The GPR and CNN models are trained with 10\%, 5\%, 1\% , 0.5\%, 0.25\%, and 0.1\% of the scenario dataset. Since, the total number of samples per scenario is 72,400 the percentages correspond to 7240, 3620, 724, 362, 181, and 72 samples, respectively. 

\subsection{FC-AE Training}
The number of neurons in each layer is shown in Table \ref{tb_FC_AE}. For the 16 $\times$  16 ADP input, the FC-AE has three encoder layers, while the for the 8$\times$8 ADP, two layers are sufficient. The table shows the number of neurons in each layer required to encoded 16 $\times$ 16 ADP to 4 $\times$ 4 encoded ADP and 8 $\times$ 8 ADP to 4 $\times$ 4 encoded ADP. Furthermore, each layer is followed by a Leaky ReLU activation function. The training model uses Mean Square Error (MSE) loss function and Adam optimizer with learning rate 0.0005. The training is performed over 1000 epochs with batch size of 256.

As a deep neural network, the FC-AE requires a large scale dataset to train the model. However, the FC-AE only needs to be trained one time after which it can be used on other scenarios with similar ADP characteristics. The more diverse the training dataset, the more robust the FC-AE to new scenarios. The dataset may be collected at a single environment or it may be sampled from multiple environments. 

\begin{table}[h]
\vspace{-5mm}
 \begin{center}
\caption{\small{FC-AE Architecture}}
\scalebox{0.9}{
\begin{tabular}{ |c|c|c| } 
 \hline
 \textbf{Layer} & \textbf{ADP 16x16}  &  \textbf{ADP 8x8} \\ 
 \hline
 Encoded Layer 1  &  256  & 64   \\ 
  \hline
Encoded Layer 2 & 144  & 36  \\ 
  \hline
Encoded Layer 3 & 64  & -  \\ 
  \hline
Bottle Neck Layer  & 16 & 16  \\ 
 \hline
Decoded Layer 1 & 64  & 36  \\
 \hline
 Decoded Layer 2& 144  & 64  \\
 \hline
  Decoded Layer 3 & 256  & -  \\
 \hline
\end{tabular}}
\vspace{-10mm}
 \label{tb_FC_AE}
\end{center}  
\end{table}

To test the performance of the FC-AE on multiple scenarios, the model is trained on Scenario S1 and tested on the remaining five scenarios. The performance on the FC-AE is measured in how well the FC-AE decodes the ADP. The similarity between the original input ADP and the decoded ADP is measured using 
similarity $\mathscr{S}(\boldsymbol{A},\boldsymbol{A_d})$ defined in \eqref{eq_sim}, where higher values of $\mathscr{S}$ imply more similarity between the two ADPs and $\mathscr{S}(\boldsymbol{A},\boldsymbol{A_d})=1$ means they are identical.  Table \ref{tb_similarity} shows the average similarity of the FC-AE model that encodes 16$\times$16 ADP to 4$\times$4 ADP. The FC-AE is first trained with S1 dataset and tested on the remaining scenario dataset. The similarity results for each scenario are shown in the second column. Scenarios S2-S3 perform well with high  similarities, but the similarity for scenarios S4-S6 drops significantly. Next, the model is retrained with S1 and S4 datasets and tested on the remaining scenarios. An improvement in similarity is observed for all scenarios. The most significant improvement is on S4, which is expected as S4 is included in the training dataset. Excluding S4, the improvement in similarity is between 1\% and 4\%. This demonstrates that the FC-AE model can be improved to be more robust to different scenarios given a large enough and diverse training dataset. 
Similarly, Table \ref{tb_similarity_2} shows the similarity performance for the FC-AE that encodes 8$\times$8 ADP to 4$\times$4 ADP.  Overall, this model performs better than the 16$\times$16 ADP since the compression ratio is smaller. Furthermore, the improvement after adding S4 to the training dataset is between 2\% and 4\%. 

\begin{singlespace}
\begin{table}[H]
\vspace{-5mm}
 \begin{center}
\caption{\small{Decoding Similarity for FC-AE that encodes 16x16 ADP to a 4x4 encoded ADP.}}
\scalebox{0.9}{
\begin{tabular}{ |c|c|c| } 
 \hline
 Scenario & \begin{tabular}{@{}c@{}}Average Similarity $\mathscr{S}$ \\ trained with S1 \end{tabular} & \begin{tabular}{@{}c@{}}Average Similarity $\mathscr{S}$  \\ trained with S1+S4  \end{tabular}\\
 \hline
S1 & 0.92 &0.92\\
 \hline
S2 & 0.91 &0.92\\
 \hline
S3 & 0.91 &0.92\\
 \hline
S4 & 0.66 & 0.96\\
 \hline
S5 & 0.63 & 0.67\\
 \hline
S6 & 0.77 & 0.78\\
\hline
\end{tabular}
}
\vspace{-10mm}
 \label{tb_similarity}
\end{center}  
\end{table} 
\end{singlespace}

\begin{singlespace}
\begin{table}[H]
\vspace{-5mm}
 \begin{center}
\caption{\small{Decoding Similarity for FC-AE that encodes 8x8 ADP to 4x4 encoded ADP.}}
\scalebox{0.9}{
\begin{tabular}{ |c|c|c| } 
 \hline
 Scenario & \begin{tabular}{@{}c@{}}Average Similarity $\mathscr{S}$ \\ trained with S1 \end{tabular} & \begin{tabular}{@{}c@{}}Average Similarity $\mathscr{S}$  \\ trained with S1+S4  \end{tabular}\\
 \hline
S1 & 0.94 & 0.96\\
 \hline
S2 & 0.94 & 0.96\\
 \hline
S3 &  0.94 & 0.97\\
 \hline
S4 & 0.87 & 0.89\\
 \hline
S5 & 0.84 & 0.87\\
 \hline
S6 & 0.92 & 0.95\\
\hline
\end{tabular}
}
\vspace{-10mm}
 \label{tb_similarity_2}
\end{center}  
\end{table} 
\end{singlespace}

Fig. \ref{fig_dADP} shows an example of the original ADP (left) and the decoded ADP (right) that have similarity of 0.95. The FC-AE is not completely lossless, which means that there will be noise in the decoded ADP.  However, the defining features of the ADP which are the strong multiplath components are preserved. Therefore, the information about the channel contained in the ADP as well as the uniqueness of the ADP remain preserved. 

 \begin{figure}[htbp]
 \vspace{-5mm}
    \centering
    \includegraphics[width=3.8in,height=1.4 in]{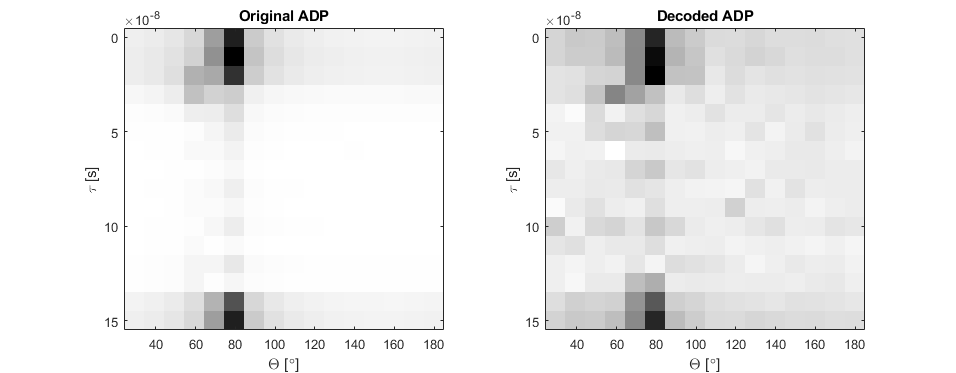}
    \caption {\small {Comparison between original input ADP and the decoded output ADP with similarity of 0.95.}}
    \vspace{-10mm}
    \label{fig_dADP}
\end{figure}

\subsection{GPR Training}

The GPR model is designed in MATLAB 2019b using the \textit{fitgpr} function \cite{matlab}. The \textit{fitgpr} function  automatically tunes the parameters and selects the best kernel function. The model uses \textit{bayesopt}, which is a Bayesian optimizer. Furthermore, model uses five-fold cross-validation loss. The training model tries to minimize the Objective Function (OF) \cite{KHUWAILEH20201807} 
\begin{equation}
  \textrm{OF}= \log(1+ \textrm{cross-validation loss}).
   \label{eq_OF}
\end{equation}
Each kernel is optimized with respect to the kernel hyper-parameters $\boldsymbol{\Theta}$ which include the kernel scale $l$ and standard noise deviation $\sigma$.   
In the section, we demonstrate the GPR optimization for a $4 \times 4$ ADP dataset. While this section specifically addresses GPR optimization for $4 \times 4$ ADP dataset, the same optimization procedure is repeated for other ADP sizes and for encoded ADPs. 

Fig. \ref{fig_GPRTrain} illustrates the optimization for the squared exponential kernel. The OF on the z-axis is optimized with respect to kernel scale $l$ and the standard deviation of noise  $\sigma$. The model is iterated over multiple $l$ and $\sigma$ combinations until the OF function converges or the algorithm reaches the maximum number of iterations of 20. Each iteration is represented with a blue dot (observed point) and the green dot represents the iteration with the smallest OF value (minimum feasible OF value found during training). The mean model surface is generated by interpolating between the blue dots. 
This process is repeated for the remaining four kernel types in Table \ref{tb_covfun} and an OF function model is created for each one of them. The minimum OF for each kernel type is summarized in Table \ref{tb_OF}. The table lists the kernel types evaluated, the minimum estimated OF value, and the values of the associated hyper-parameters ($l$ and $\sigma$) when the OF is minimum.  The results in the table show that the minimum OF occurs for the squared exponential kernel with $l = 2.4*10^{-3}$ and $\sigma = 5.4$.

  \begin{figure}[h]
 \vspace{-5mm}
    \centering
    \includegraphics[width=4.5in,height=3.0 in]{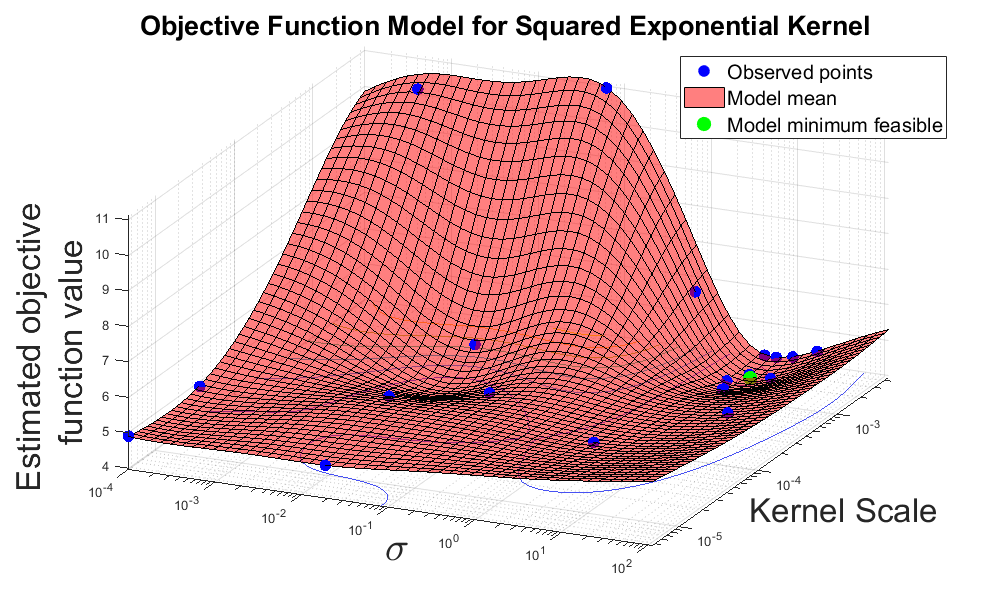}
    \caption {\small {The OF model for the squared exponential kernel. The OF on the z-axis is optimized with respect to 
   standard deviation of noise $\sigma$ and the kernel scale $l$.
    The OF is estimated at each $l$ and $\sigma$ combination depicted with a blue dot (observed point). The green dot represents the location of the minimum estimated OF. The red surface is formed by interpolation between the observed points.}  }

    \vspace{-5mm}
    \label{fig_GPRTrain}
\end{figure}

\begin{table}[h]
\vspace{-5mm}
 \begin{center}
\caption{\small{Results for Minimum OF for each kernel type. }
}
\scalebox{1.0}{%
\begin{tabular}{ |c|c|c|c|} 
 \hline
 \textbf{ Kernel } & \textbf{ Minimum OF }  &\textbf{Kernel Scale $l$} & \textbf{$\sigma$} \\
 \hline
    Squared Exponential & \textbf{3.9637} & $2.4*10^{-3}$ & 5.4 \\
     \hline
 Exponential & 4.0049 & $1.7*10^{-3}$ & $1.1*10^{-4}$ \\
  \hline
 Rational Quadratic & 4.0627 & $2.5*10^{-3}$ & $7.6$\\
 \hline
 Matern32 & 4.1482 & $1.8*10^{-4}$ & $1.3*10^{-3}$\\
 \hline
 Matern52 & 4.0166 &$2.6*10^{-3}$ & $5.3$ \\
\hline
\end{tabular}}
 \label{tb_OF}
 \vspace{-10mm}
\end{center}  
\end{table}

\subsection{CNN Training}
The CNN is designed as an $L$ layer regression network. The kernel size and number of filters for each layer are listed in Table \ref{tb_CNN}. Furthermore, the CNN uses ReLU for the activation function and each covolutional layer is followed by a maxpool layer with kernel size $2\times2$. There is also a fully connected layer at the output. The CNN is trained over 1000 epoch with a batch size of 100.

\begin{singlespace}
\begin{table}[h]
\vspace{-5mm}
 \begin{center}
\caption{\small{CNN setup for each of the ADP input sizes}}
\scalebox{1.0}{
\begin{tabular}{ |c|c|c|c| } 
 \hline
 \textbf{Layer} & \begin{tabular}{@{}c@{}}\textbf{kernel x filter} \\\textbf{ADP 16x16} \end{tabular} & \begin{tabular}{@{}c@{}}\textbf{kernel x filter} \\ \textbf{ADP 8x8} \end{tabular} & \begin{tabular}{@{}c@{}}\textbf{kernel x filter} \\ \textbf{ADP 4x4} \end{tabular} \\
 \hline
 1  &  $8\times 8 \times 8$  & $7 \times 7 \times 16$ &  $5 \times 5 \times 32$   \\ 
  \hline
 2 & $7 \times 7 \times 16$  &  $5 \times 5 \times 32$  & $3 \times 3 \times 64$ \\ 
  \hline
3  & $5 \times 5 \times 32$ & $3 \times 3 \times 64$  & \\ 
 \hline
4  & $3 \times 3 \times 64$  & &  \\
 \hline
\end{tabular}}
 \label{tb_CNN}
 \vspace{-10mm}
\end{center}  
\end{table}
\end{singlespace}

\section{Results and Discussion}
\label{Results}

\subsection{Localization Performance on Scenario S2}
This section analyzes the FC-AE-GPR performance on the scenario S2. The FC-AE-GPR is compared with the CNN regression model and the GPR model. The results are shown in Table \ref{tb_results_s1}. The first column shows the size of the ADP input. 
{The purpose of the FC-AE is to reduce the input size before training the GPR model.} Both 16$\times$16 and 8$\times$8 ADP inputs are compressed to 4$\times$4. For 4$\times$4 ADP, the input size is already small, so no further compression is performed. In this case, only the CNN and GPR performances are compared. Moreover, columns 3-8 show what percentage of the total dataset was used for training the model. In these columns, the performance of the models is compared with respect to the size of the training dataset.  Furthermore, since the training dataset is small, the performance of the model highly depends on the training samples selected. Therefore, the simulations are repeated 50 times with different training samples for each simulation and the results are reported as the mean RMSE in meters over 50 trails.

When the number of training data points is larger, all three models perform well for 16$\times$16 and 8$\times$8 ADP input. However, as the training dataset becomes smaller, the GPR and FC-AE-GPR outperform the CNN. Furthermore, the CNN error is initially changing slowly until it reaches a "breaking point" after which the CNN error escalates quickly. This breaking point occurs somewhere between 5\% and 1\% for 16$\times$16 ADP, and between 0.25\% and 0.1\% for 8$\times$8 ADP.
In some instances, the FC-AE-GPR performance experiences a slight decline compared to the GPR at the expense of significantly reducing the computational speed. However, the FC-AE-GPR still outperforms the CNN for small size training datasets. The performance of the FC-AE-GPR may be further improved by improving the  diversity of the FC-AE. 

Regarding the size of ADP input, the CNN performs well when the ADP dimensions are larger, but its performance starts to degrade for smaller ADP sizes. On the other hand, the GPR model is much more robust to small ADP inputs and the performance degrades at a much slower rate.  Specifically, for ADP 4$\times$4, the CNN completely fails with RMSE of more than 15.0 m, while the GPR can still perform localization with reduced accuracy. For example, the CNN mean RMSE is 15.07 m for training size of 10\%, while the GPR mean RMSE for the same training size is only 4.27 m. 

\begin{singlespace}
\begin{table}[h]
  \centering
  \caption{\small{Localization error evaluated for different ADP dimensions in Scenario S2.}}
    \begin{tabular}{cccccccc}
   \cmidrule{3-8}        
 \multicolumn{2}{c}{mean RMSE (m)}    &    
\multicolumn{6}{c}{Percentage of Total Dataset Used for Training} \\
\cmidrule{3-8}        ADP size         &  Type  & 10.0\% & 5.0\% & 1.0\% & 0.5\% & 0.25\% & 0.1\% \\
\hline
 
    \multirow{2}[2]{*}{16 $\times$ 16} 
                    & CNN  & 0.56 &0.98 &15.89 &15.33 & 16.84 &57.36\\
                    & GPR & 0.60 &0.98 & 2.56 & 3.91 & 5.43 &7.52\\
                    & FC-AE-GPR & 0.65 &1.03 & 2.89 &4.11 & 5.81 & 8.76  \\

\cmidrule{1-8}           
    \multirow{2}[2]{*}{8 $\times$ 8} 
                    & CNN & 1.79 & 1.99 &4.59 &6.36 &7.83 & 42.17 \\
                    & GPR & 0.69 &1.76 &4.20 &5.77 &7.16 & 10.56\\
                    & FC-AE-GPR &
                    0.71 &1.85 &4.24 &5.78 &7.03 &12.81 \\
         
 \cmidrule{1-8}           
    \multirow{2}[2]{*}{4 $\times$ 4} 
                  & CNN  & 15.07 &16.03 &16.55 &17.29 &19.29 &75.51\\
                  & GP & 4.27 &5.05 &7.50 &8.81 & 10.8 & 13.4\\
\hline

    \end{tabular}
  \label{tb_results_s1}
  \vspace{-10mm}
\end{table}
\end{singlespace}

\subsection{Offline Training Time}
One of the main concerns of GPR model design is the computational time involved in training the model. The computational complexity requirement for GPR optimization is  $O(n^3)$, where $n$ is the number of training samples. The cubic complexity comes from the inversion and determinant of the kernel matrix $\mathbf{K}$ in \eqref{eq_nlml}. Furthermore, the computation of kernel functions $k_{ij}(\mathbf{x}_j,\mathbf{x}_j)$ that form the kernel matrix $\mathbf{K}$ depend on the size of the input $\mathbf{x}$ and have a computational complexity of  $O(d^2)$, where $d$ is the length of the input vector $\mathbf{x}$  (vectorized ADP). The squared complexity comes from the vector multiplication in \eqref{eq_cov_r}.

The graph in Fig. \ref{fig_opt_time} shows how the GPR optimization time relates to the number of training samples ($n$) and the dimension of the input ($d$). As the number of training samples increases ten times from 72 to 724 samples, the computation increases by approximately two orders of magnitude regardless of the ADP input size. In addition, as the size of the ADP increase from $4^2$ to $8^2$ and then to $16^2$, the optimization time increases by one and two orders of magnitude, respectively. Therefore, by encoding the $8^2$ and $16^2$ ADP inputs to $4^2$ input, the optimization time significantly reduced. 
 
 \begin{figure}[h]
 \vspace{-5mm}
    \centering
    \includegraphics[width=3.4in,height=2.4 in]{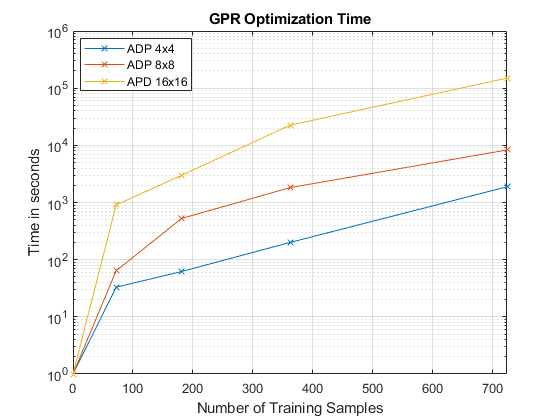}
    \caption {\small{ Hyper-parameter optimization time of the GPR model   }}
    \vspace{-10mm}
    \label{fig_opt_time}
\end{figure}

\subsection{Localization Performance on New Scenarios}
In this section, the independence of the FC-AE is evaluated with respect to the localization error. The FC-AE is trained with scenario S1 dataset and then the localization performance is evaluated for  S1-S5. Scenario S6 is intentionally omitted in this analysis as the grid size of this dataset is different from the first five scenarios and the localization performance would not be a direct comparison. 
The mean RMSE (in meters) over the 50 trails is shown in Table \ref{tb_results_multipleS}. As expected, there is a strong correlation between the similarity reported in the second column in Table \ref{tb_similarity} and the localization framework performance. When the similarity is high ($>0.8$), the FC-AE-GPR performs with high accuracy as is the case for S1-S3. However, as the similarity decreases, the ADPs are not encoded well resulting in a poor performance. Poor FC-AE performance eliminates the uniqueness of the ADP input which not only impacts the testing sample, but it also impacts the training of the GPR model. The GPR model cannot be trained properly if the ADP samples are not unique to the user's location. The similarity of S5 is less than the similarities of S1-S3,  which is reflected in the performance, where the RMSE of S5 is higher than the RMSE for S1-S3. However, S5 can still perform localization unlike S4. The similarity of S4 is very low as shown in Table \ref{tb_similarity}, causing the GPR training model to be unable to accurately estimate the location using the encoded ADPs, resulting in a significant level of error. 


\begin{singlespace}
\begin{table}[h]
  \vspace{-5 mm}
  \centering
  \caption{ \small{Localization RMSE in meters for FC-AE-GPR model for 16 x 16 ADP encoded to 4 x 4.}}
  \begin{tabular}{c|cccccc}
\cmidrule{1-7}        
 \multicolumn{2}{c}{   }  &
\multicolumn{5}{c}{Percentage of Total Dataset Used for Training} \\
\hline
Scenario & 10.0\% & 5.0\% & 1.0\% & 0.5\% & 0.25\% & 0.1\% \\
\hline
\hline
S1  & 0.65& 1.03 & 2.89 & 4.31 &6.20 & 9.14\\      
S2  &0.67 & 1.07 &2.92&4.11 &6.01 & 8.97 \\                 
S3  & 0.62 &0.98 &2.72 &4.04 &5.81& 8.76\\                 
S4  & 38.79 & 38.85 & 39.01 & 39.13 & 39.13 & 39.31\\
S5  & 0.75 &1.15 &3.59 &5.46 &7.82 & 11.38\\               
\hline
\end{tabular}
  \label{tb_results_multipleS}
  \vspace{-10 mm}
\end{table}
\end{singlespace}

\section{Conclusion}
\label{Conclusion}
This paper presented a fingerprinting-based localization framework, referred to as FC-AE-GPR, for single-site massive MIMO systems. The FC-AE-GPR is a novel framework that combines Fully-Connected Auto-Encoder with Gaussian Process Regression models. The FC-AE-GPR framework is designed for small-scale training dataset. The ability to train on a small-scale dataset is accomplished by using the GPR model, while the computational complexity is reduced by using an FC-AE to compress the size of the ADP input prior to training GPR model.
Since the FC-AE is scenario independent, adapting the FC-AE-GPR to a new, unseen scenarios requires only a small training dataset to retrain the GPR model.  
The simulation results show that the both GPR and FC-AE-GPR outperform the state-of-the art CNN regression model for small training dataset. Furthermore, the FC-AE improves the GPR model by significantly reducing the computational time of the model which is the main limitation of GPR models. 



\bibliographystyle{IEEEbib}
\bibliography{references}
\end{document}